\documentclass[aps,10pt,pra,twocolumn,showpacs]{revtex4-1}
\usepackage{graphicx}
\usepackage{url}

\usepackage[euler]{textgreek}

\begin{document}

\title{Integrated-photonic characterization of single-photon detectors for use in neuromorphic synapses}

\author{Sonia M. Buckley}\email[]{sonia.buckley@nist.gov}
\author{Alexander N. Tait}
\author{Jeffrey Chiles}
\author{Adam N. McCaughan}
\author{Saeed Khan}
\author{Richard P. Mirin}
\author{Sae Woo Nam}
\author{Jeffrey M. Shainline}

\affiliation{National Institute of Standards and Technology, 325 Broadway, Boulder CO 80305}

\begin{abstract}
Large-scale spiking neural networks have been designed based on superconducting-nanowire single-photon detectors (SNSPDs) as receiver elements for photonic communication between artificial neurons and synapses. Such large-scale artificial neural networks will require thousands of waveguide-integrated SNSPD devices on a single chip. Efficient waveguide-coupled SNSPDs and high-throughput methods for their characterization at this scale are very challenging. Here, we design, fabricate, and measure SNSPDs that are compatible with these large-scale networks. We demonstrate integrated-photonic circuits to simultaneously characterize many waveguide-coupled SNSPDs with a single fiber input. We achieve up to 15 waveguide-coupled SNSPDs in a single integrated-photonic circuit and a total of 49 waveguide SNSPDs with a detection plateau out of 49 tested. To our knowledge, this is the largest number of SNSPDs integrated in a photonic circuit to date. We employ several types of photonic circuits to enable rapid and reliable characterization of detector performance. We further demonstrate several important synaptic functions of these detectors. These include a binary response in the detectors with average incident photon numbers ranging from less than $10^{-3}$ to greater than $10$, indicating that synaptic responses based on these detectors will be independent of the number of incident photons in a communication pulse. Such a binary response is ideal for communication in neural systems. We further demonstrate that the response has a linear dependence of output current pulse height on bias current with up to a factor of 1.7 tunability in pulse height, a function that cannot be obtained without a detection plateau.\end{abstract}

\maketitle

\section{Introduction}
\label{sec.introduction}

A defining feature of neural systems is the presence of highly interconnected neurons making thousands of synaptic connections that are often spatially distant. Many electronic circuits based on semiconductors \cite{Mead1989,voma2007,inli2011,cryu2012,pfgr2013,cage2013,bega2014,Chicca2014,Furber2016} and superconductors \cite{hago1991,hiak1991,koko2005,hias2006,hias2007,crsc2010,sele2017,scdo2018} have been explored to achieve synaptic and neuronal computations, but these circuits attempt to perform communication as well as computation in the electronic domain.  It has been argued extensively that artificial neural systems will benefit from employing photonic rather than electronic communication to achieve this high fan-out \cite{abps1987,delima2017,Shainline2019,Moughames2020}. A number of implementations of neural systems employing light for communication have been explored \cite{psfa1985,Caulfield1989a,Tait2014,Shen2017,Feldmann2019}, but these typically do not include the same diversity of circuits for learning and computation, and require much higher energy sources and detectors. We have proposed an approach that combines photonic circuits for communication with superconducting electronics for computation \cite{Shainline2017a,Shainline2018,Shainline2019,Shainline2020}. In this approach an LED and transmitter circuit mimics the firing operation of the neuron, and the phase transition of an SNSPD converts the optical pulse to a binary electrical pulse, in analogy with the electrical to chemical conversion that occurs at the synaptic cleft. Synaptic, dendritic, and neuronal computations make use of the inherent thresholding nonlinearity of Josephson junctions.

While superconducting optoelectronic hardware has been proposed as a means to achieve large-scale neural systems, significant experimental progress is required to prove the practical viability. Before the full neuronal circuit can be realized, individual subcomponents must be fabricated and tested, so that each is performing as needed before integration. There are four main device subcomponents under investigation for future integration \cite{Buckley2020a}: waveguide-coupled SNSPDs (this work), waveguide-coupled LEDs \cite{Buckley2017}, amplifier devices \cite{McCaughan2019}, and SNSPDs integrated with Josephson junctions. While waveguide-coupled SNSPDs have been demonstrated previously, neural networks with modest numbers of neurons will require thousands of SNSPD devices integrated with waveguides, light sources, and superconducting electronics. Therefore one of the major contributions of this work is the high yield of the detectors on single-mode silicon waveguides, allowing integration of large numbers (up to fifteen in this work) of SNSPDs with saturating internal efficiency in a single photonic circuit. 

There are many aspects of an SNSPD that must be considered when designing a synapse, including critical current, inductance, device area, operating temperature, detection efficiency, detection plateau width, device yield, and superconductor material. All of these properties are interconnected, and therefore they must be optimized self-consistently in the system context. Fabrication and testing of waveguide-SNSPDs with the most likely parameters to be used in a system integration is therefore a necessary precursor to building neuronal circuits. In Section \ref{sec.cryo}, we focus on experimental characterization and measurement techniques of waveguide-coupled SNSPDs with parameters similar to those expected to be utilized in the full neuronal circuit demonstrations. Because synapses are the most numerous elements in neural systems, the ability to rapidly characterize many SNSPDs in the integrated-photonic environment in which they will be utilized is necessary both initially and for any further system optimization. Therefore both the characterized parameters and measurement techniques are highly relevant for the synaptic application. 

The requirement for saturated internal efficiency in the definition of the yield of SNSPDs has often been overlooked in the literature, despite being of critical importance for most applications due to its direct impact on detection efficiency. It is of particular importance in this work for implementations of synaptic function. In Section \ref{sec.synapses} we show how technological improvements that lead to high-yield detectors directly enable a technique for setting analog synaptic weights, an exciting space in applied physics. While synaptic weights may be either digital or analog, signaling between neurons is binary. Binary signaling is an important technique to reduce noise in communication from a neuron to its synaptic connections, and the effectiveness of communication in neural systems is due in part to the binary nature of action potentials. Binary communication is therefore a key feature of the superconducting opto-electronic hardware platform, and we demonstrate the conversion of photonic pulses with intensities varying by up to four orders of magnitude to binary electrical pulses. Finally in Section \ref{sec.discussion} we discuss the relevance of this work in a broader context.

\section{Integrated photonic characterization of single-photon detectors}
\label{sec.cryo}

Waveguide-integrated SNSPDs are being pursued for a variety of applications in integrated quantum optics, metrology and large-scale neural systems. Several review articles have been written on the subject \cite{Dietrich2016, Ferrari2018, Elshaari2020}, all of which present positive prospects for the future of systems integrating large numbers of waveguide-coupled SNSPDs. However, it is also clear that further work is needed to improve the yield of high-quality, high-performance detectors. Refs. \citenum{Elshaari2020} and \citenum{Najafi2015} suggest that a `pick-and-place' technique will allow prescreening of high-quality detectors before individual placement in photonic circuits, while Ref. \citenum{Ferrari2018} predicts that once SNSPDs are fabricated in commercial foundries yield issues will be eliminated. 

In this work we follow most closely the approach suggested in Ref. \citenum{Dietrich2016} to realizing high-yield detectors, by using the amorphous material WSi. To date, this is the material that has been shown to yield the largest number of detectors with a broad plateau \cite{Marsili2013}, although other amorphous materials with higher transition temperature are strong candidates as well \cite{Verma2015} and may ultimately prove more promising due to reduced cryogenic complexity. WSi nanowire detectors have been demonstrated in kilopixel arrays with yields greater than $99$\% \cite{Wollman2019}. 

Recent work has shown that detection plateaus can be achieved in much wider wires than previously thought possible \cite{Korneeva2018,Chiles2020}. We use the silicon-rich WSi developed in Ref. \citenum{Chiles2020}, which allows straightforward fabrication and patterning with photolithography while giving a broad detection plateau. The ability to photolithographically pattern SNSPDs greatly reduces the fabrication time and complexity compared to electron beam (e-beam) lithography, which has much longer write times and is a more manual process. Every e-beam step eliminated significantly decreases the fabrication turn-around time. For these reasons, we study WSi nanowires patterned with photolithography in this work. It is the use of this material that has enabled the integration of large numbers of detectors with saturated internal efficiency in a single device, a task that has previously proved challenging \cite{Najafi2015,Gourgues2019}. In this work we have exceeded the number of working waveguide-coupled superconducting detectors in any prior single integrated device \cite{Najafi2015, Schuck2016, Khasminskaya2016, Kahl2017, Buckley2017, Kovalyuk2018, Ferrari2018, Gourgues2019}, yielding the most complex integrated-photonic circuits utilizing SNSPDs published to date.

The ability to fabricate devices with large numbers of waveguide-coupled SNSPDs leads to challenges in testing. One major challenge in the measurement of waveguide-integrated SNSPDs is to couple light from an optical fiber into an on-chip waveguide. Due to variations in fiber coupling efficiency and the complexity of the fiber coupling/packaging procedure, it is helpful to make as many measurements as possible using a single fiber input, particularly when one wishes to compare the efficiency of different detector designs. We therefore use waveguides and beamsplitters to route light to different devices from a single fiber input. This is in contrast to the previous work on the characterization of waveguide-coupled SNSPDs, in which characterization devices typically only include a single SNSPD per input grating coupler \cite{Sprengers2011, Pernice2012, Kovalyuk2013, Schuck2013, Rath2015, Kahl2015, Kahl2016, Ferrari2019, Haussler2020}. The integrated devices in this work therefore allow rapid characterization without realignment and reliance on grating or coupler reproducibility. 

In Sec.\,\ref{sec.trees} we describe the integrated-photonic tree device, which evenly divides input light between seven detectors and can be used to compare different detector designs. In Sec.\,\ref{sec.hidras} we describe the HiDRA. The full integrated circuits consist of up to 15 detectors, integrated with up to 18 beamsplitters/beamtaps and several centimeters of waveguides. One promising light source used in the proposed superconducting spiking neural systems operate at wavelengths of around 1.22\,\textmu m \cite{Shainline2017a, Buckley2017}. Therefore, we design and demonstrate all devices for operation at both the standard 1.55\,\textmu m wavelength used for telecommunications and most quantum optics experiments, as well as 1.22\,\textmu m.

\subsection{Branching tree structures}
\label{sec.trees}
\begin{figure}
\includegraphics[width=8.6cm]{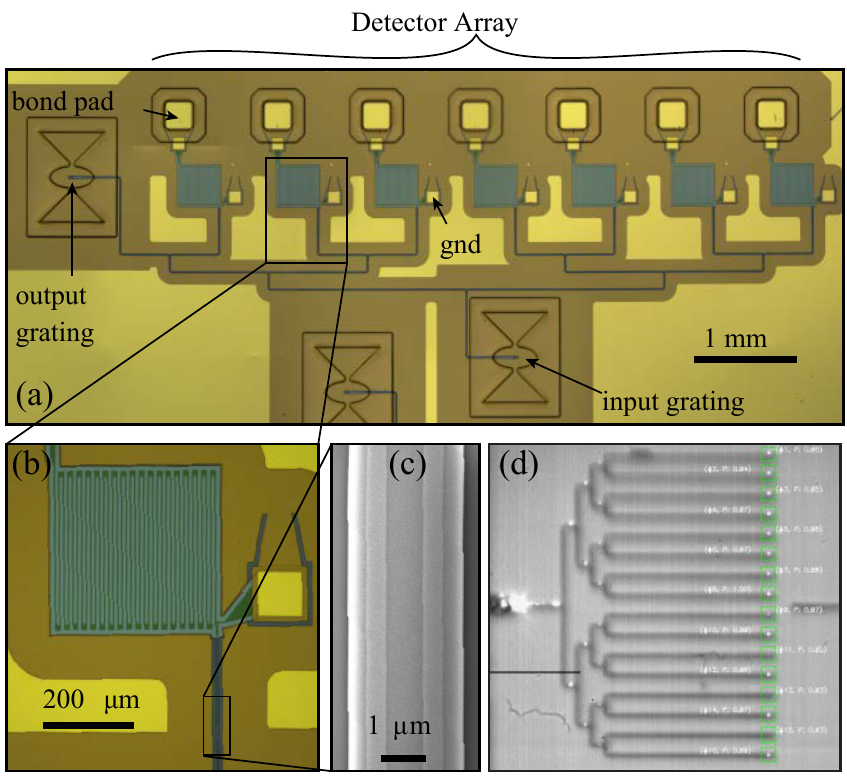}
\caption{\label{fig.tree_images} (a) An optical microscope image of the tree device. (b) Zoom-in view of a waveguide-integrated detector with associated meander inductor. (c) Scanning electron micrograph of the nanowire on the silicon waveguide. (d) Infrared camera image showing the passive tree device and the laser input. The green boxes are the automated detection of the output ports, and the white text is automated output from the program indicating the coordinates and intensity of the output.}
\end{figure}

\begin{figure}
\includegraphics[width=8.6cm]{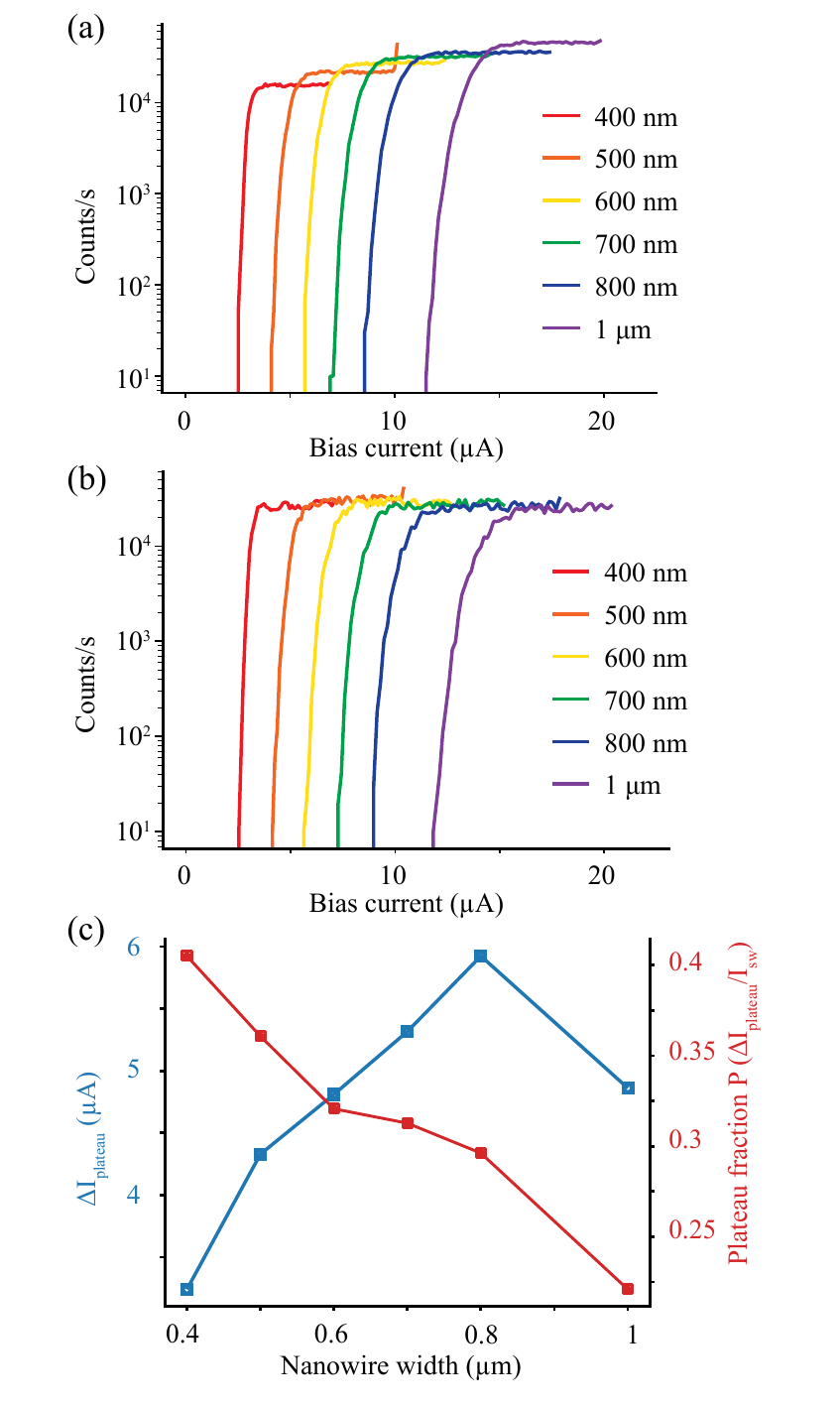}
\caption{\label{fig.tree_data} (a) Count rate versus bias current for six different SNSPD widths with uniform illumination of the detectors. (b) Count rate versus bias current for six different SNSPD widths with waveguide-coupled light. (c) The width of the plateau versus the nanowire width.}
\end{figure}

The `branching tree' structure provides a method for evenly distributing light to a number of different detectors. This allows comparison of the performance of several detectors under uniform waveguide-coupled illumination. A microscope image of such a device is shown in Fig.\,\ref{fig.tree_images}(a). Light from a single fiber input is split evenly by a series of beamsplitters and routed to seven detectors and an output grating. The output grating is used for fiber alignment during packaging at room temperature. The path lengths from the input grating to all detectors are equal. A zoom-in of a detector is shown in Fig.\,\ref{fig.tree_images}(b), and a scanning electron micrograph of an SNSPD detector on a waveguide is shown in Fig.\,\ref{fig.tree_images}(c). We have also implemented a version of this device for room-temperature characterization. The room-temperature structures include an output grating after the hairpin device, as well as output reference gratings for normalization of the output power. Absorption of the hairpins at room temperature can then be compared to measurements of hairpins at cryogenic temperature. Figure \ref{fig.tree_images}(d) shows a microscope image of such a tree structure used for room temperature characterization. The measurements described in this section were all performed at 1.55\,\textmu m on devices designed for operation at this wavelength. Once cooled, the current-voltage curves for each SNSPD detector were measured. From these data, the critical current of the standard detectors (200\,\textmu m long, 500\,nm wide) was found to be 10\,\textmu A. 

With the light evenly divided, comparisons can be made between detectors with different geometric parameters.  A layout error caused one of the seven detectors on each tree to fail, so each tree in this work compares six detectors. For the first tree, the width of the SNSPDs was varied from 400\,nm to 1.5\,\textmu m, keeping the length of the nanowire constant (see Fig.\,\ref{fig.fabrication}(c)). The results are shown in Fig.\,\ref{fig.tree_data}. Detector count rate is plotted in Fig.\,\ref{fig.tree_data}(a) when the detectors are flood-illuminated from above and in Fig.\,\ref{fig.tree_data}(b) when the detectors are illuminated through the waveguide tree structure. In both the uniform and waveguide illuminated plots the count rate is normalized to the count rate on the brightest detector. The detectors under uniform illumination required approximately 50\,dB more light for the same count rate as the waveguide-coupled devices. A plateau region where the count rate on the detector is independent of bias current is observed for every width measured, as expected based on the results in Ref.\,\onlinecite{Chiles2020}. This plateau region is the desired bias current range of operation. The narrower the wire, the steeper the transition between the onset of detection and the plateau region. This has also been characterized for this and different film compositions for uniformly illuminated detectors in Ref.\,\onlinecite{Chiles2020}. 

We further use the tree structure to investigate the height and width of the plateau region. The height of these curves quantifies the relative detection efficiency of wires with differing widths, while the width of the plateau quantifies the range of bias currents providing saturated internal quantum efficiency. We find very little dependence of detection efficiency on wire width (see Methods Sec. \ref{sec.tree_analysis}). Figure \ref{fig.tree_data}(c) shows the plateau width, $P$, versus the wire width $w$. The value of $P$ is calculated as the ratio of the width of the region between the value of $I_{bias}$ at 0.9 of the maximum count rate and the switching current $I_{sw}$ ($\Delta I_{\mathrm{plateau}}$), to the switching current $I_{sw}$: $P \equiv \Delta I_{\mathrm{plateau}}/I_{\mathrm{sw}}$. For large-scale systems, the definition of yield should include this measure, and therefore we will not include any SNSPDs with $P < 0.1$ in counting the number of working devices for the yield calculation. In fact, the 500\,nm wide standard detectors used in this work had a typical value of $P$ much larger than this, with $P$ between 0.3 and 0.4. We also include the length of the plateau region as an absolute current value on the righthand axis of Fig. \ref{fig.tree_data} (c), as in experiments this may be more important than the fractional value $P$ due to equipment and noise limitations. 

The optimal performance depends on the application and the circuit used to read out the signal. Based on this analysis, any of the detector widths chosen would work for applications at 1.55\,\textmu m. The widest wires are the least sensitive to lithography, but have the largest area and highest energy consumption, while the narrowest wires' reduced performance suggests this is close to the resolution limit of our 365\,nm i-line lithography process.

A second tree device with nanowires of different lengths was also fabricated and measured. The nanowire length was varied from 5\,\textmu m to 200\,\textmu m, based on simulations of the nanowire absorption length (see Methods Sec. \ref{sec.photonic_design}). We find an absorption length of 160 dB/mm when measuring the detectors cryogenically at 1.55\,\textmu m  (see Methods Sec. \ref{sec.tree_analysis}), and 175\,dB/mm when measured at room temperature. The room temperature measurement technique is automated and highly parallel, and is described in Methods Sec. \ref{sec.high_dynamic_range}. Similar room-temperature/cryogenic comparison has been performed previously for NbN SNSPDs on nanophotonic silicon and diamond waveguides \cite{Pernice2012, Kahl2016}, and room temperature absorption measurements of similar devices have been performed for NbN on diamond and SiN waveguides \cite{Kovalyuk2013,Rath2015}. In all previously published experiments the detectors measured were each on a separate integrated photonic device, rather than being integrated in a single photonic circuit, as in this work. The absorption values in this work are similar to those found in diamond (50-175\,dB/mm for a single SNSPD hairpin) or silicon nitride waveguides. The absorption values are lower than the absorption coefficient (around 1000 dB/mm) found previously on silicon waveguides. This is due to 1) the thicker and wider silicon waveguides used in this work (220\,nm thick instead of 110\,nm thick) leading to a lower evanescent field, 2) the presence of the silicon nitride spacer between the waveguide and the nanowire and used as an etch stop, and 3) the thinner superconducting film used (2\,nm thick WSi versus 3.5\,nm thick NbN). These choices were made for ease of processing and would be straightforward to remedy in a more sophisticated foundry.  Discussion of room temperature characterization of tree devices at 1.22 \textmu m is in Sec.\,\ref{apx.1220}.

\subsection{High-dynamic-range detector arrays}
\label{sec.hidras}
\begin{figure}
\includegraphics[width=8.6cm]{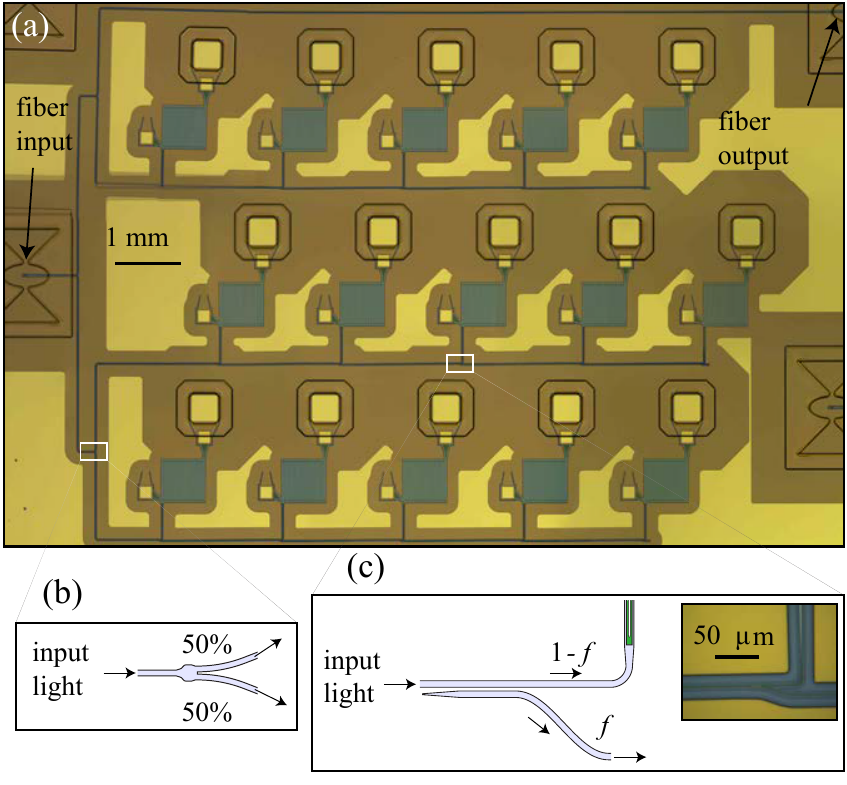}
\caption{\label{fig.hidra_images}The HiDRA device used to test the dynamic range of the integrated SNSPD platform. (a) Optical microscope image of the HiDRA device. (b) Schematic of a 50/50 beamsplitter. (c) Schematic and optical microscope image of a beamtap.}
\end{figure}

The next device examined in this study is a high-dynamic-range detector array (HiDRA). These were previously proposed and demonstrated in Ref. \citenum{Buckley2017}, but the devices did not work as predicted due to the presence of multiple waveguide modes. In this work, all waveguides are single mode. An optical microscope image of the device is shown in Fig.\,\ref{fig.hidra_images}(a). Each on-chip system consists of three HiDRAs with a single input grating and a single output grating used for fiber alignment during packaging. Light from the input grating is split evenly to four waveguides by three 50/50 beamsplitters (see Fig.\,\ref{fig.hidra_images}(b)). Three of these four branches are attached to a HiDRA, while the fourth is a straight waveguide connected to an output grating. The HiDRA device works as follows. Input light is incident on a beamtap (see Fig.\,\ref{fig.hidra_images}(c)). The beamtap drops a fraction $1-f$ of the light to an SNSPD while a fraction $f$ of the light is passed to the next beamtap. 

\begin{figure}
\includegraphics[width=8.6cm]{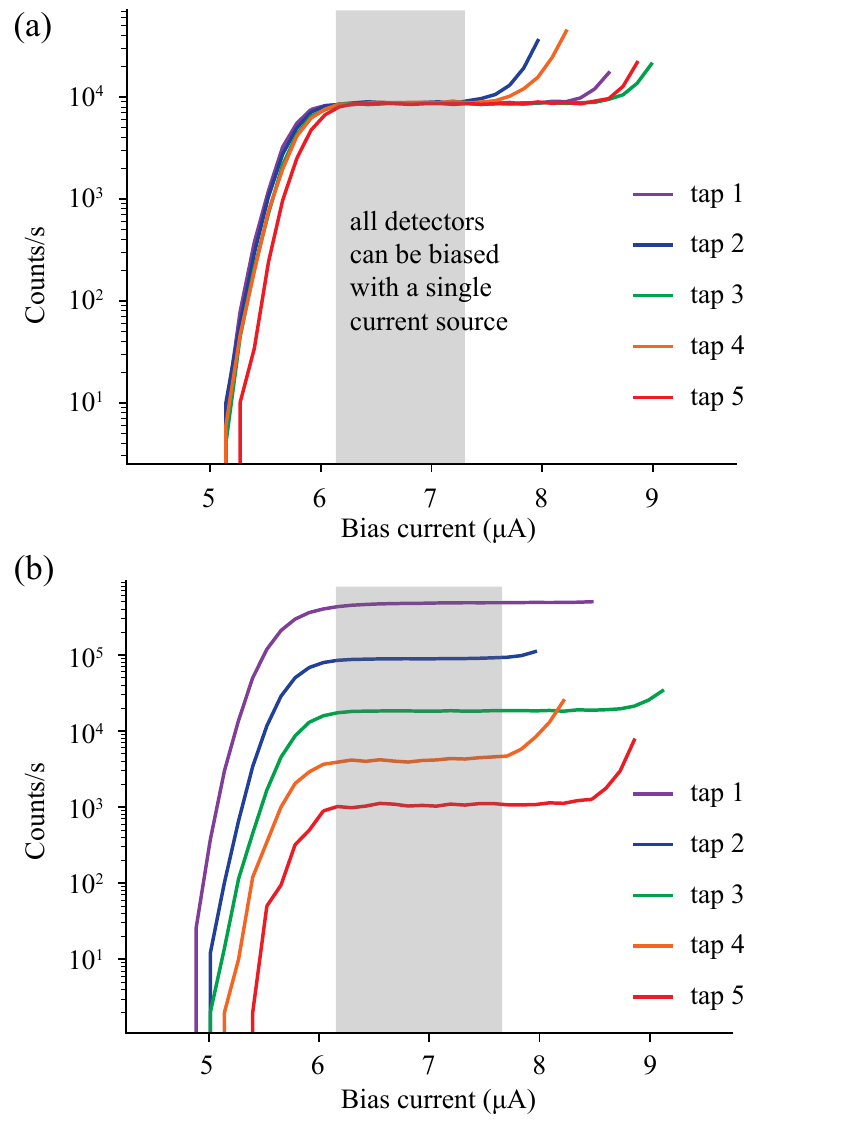}
\caption{\label{fig.hidra_data} (a) Count rate versus bias current for the five detectors in a HiDRA device when the chip is uniformly illuminated. (b) Count rate versus bias current for the same device as in part (a) with the input light waveguide coupled.}
\end{figure}

Figure \ref{fig.hidra_data} shows data acquired from such HiDRAs. Figures \ref{fig.hidra_data} (a) and (b) show counts per second versus bias current for the five detectors on a HiDRA designed to have a 10\% beamtap ($f=0.1$). Figure \ref{fig.hidra_data}(a) shows the response of the detectors when the chip is uniformly illuminated. The detectors are designed to be identical and have very similar responses to this uniform illumination. Figure \ref{fig.hidra_data}(b) shows the response when the light is waveguide-coupled, where successive detectors have significantly lower count rates. In each of Fig. \ref{fig.hidra_data} (a) and (b) the gray shaded region illustrates the range of bias currents over which the HiDRA can be biased with a single current source on all detectors (guide to the eye). The range is truncated on the left by a decrease in detection efficiency, and on the right by an increase in dark counts. The dark counts increase faster in the uniform illumination case due to heating from the higher optical powers needed in this case. The ability to bias many SNSPDs with a single input is important when scaling to hundreds of SNSPDs on-chip, as there will not be enough cryogenic I/O lines for a single individually tuned line per device as is typically used in testing experiments. We extract tap fractions of 0.26 and 0.28 from the two HiDRAs designed to have taps with $f=0.1$, with the mean value (solid black line) of 0.27. This deviation from the designed value is likely due to the the gaps coming out narrower than designed. This in combination with a slightly longer effective length due to the sine bends leading up to the beamtap could explain the discrepancy. A background light level around 30\,dB below the level of the first waveguide coupled detector on the connected HiDRA was also measured using disconnected devices. This indicates that with our current fabrication process and packaging scheme there is a 30\,dB noise floor. Previous measurements of LEDs coupled to waveguides indicated a 40\,dB noise floor \cite{Buckley2017}. Both the low fiber-coupling efficiency and high waveguide scattering loss contribute to this high background light level (see Methods for a precise characterization of these losses). As in the case of trees, `passive HiDRA' structures were fabricated for characterization at room temperature. Unlike the trees, in the case of the HiDRA structures measured at room temperature, the SNSPD has been omitted, and the structure is intended to characterize the photonic beamtap ratios. The measurement technique used for characterization of the structures is discussed in Methods Sec. \ref{sec.high_dynamic_range}. The beamtap ratio at room temperature is determined to be 0.29 for these devices, which agrees well with the cryogenic measurements described in the preceding paragraph.We also designed HiDRA to have 1\% and 0.1\% beamtaps, however, in the fabricated devices the scattered background light prevents such low tap ratios from being useful. In Methods (Section \ref{apx.1220}) we make use of the good agreement between cryogenic and room-temperature measurements for characterization at 1.22\,\textmu m. We test the detectors under uniform illumination only, and then perform measurements of the passive integrated-photonic structures at 1.22\,\textmu m to avoid the lengthy packaging and cool down processes. For the HiDRA, the noise floor was much higher for the 1.22\,\textmu m devices, at around 15\,dB, likely due to a combination of lower grating coupling efficiency (around 3\,dB lower per coupler) and higher propagation loss (14\,dB/cm higher loss). See Methods for details. 

\section{Considerations for synaptic operation}
\label{sec.synapses}
\begin{figure}
\includegraphics[width=8.6cm]{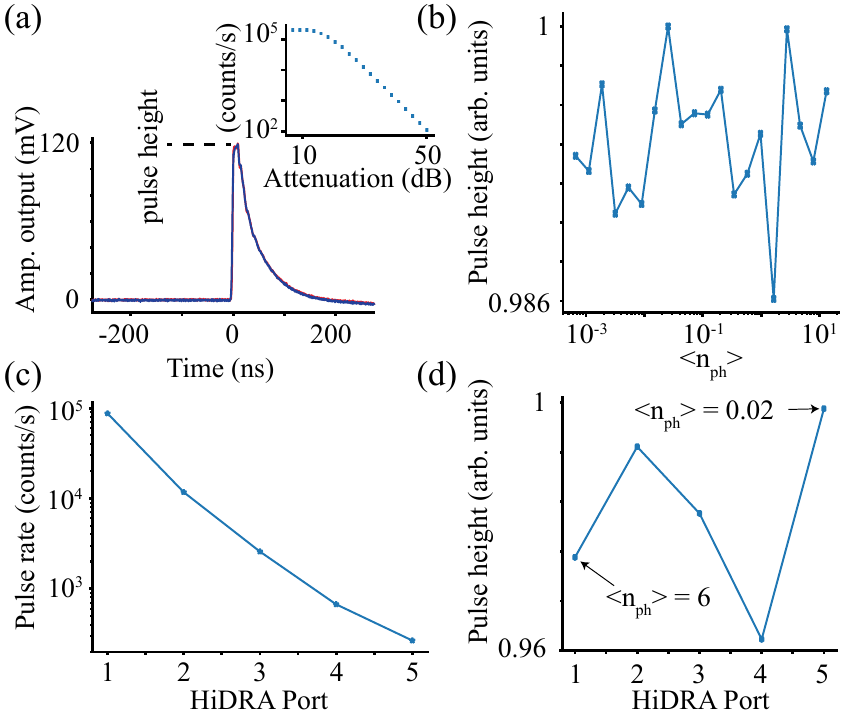}
\caption{\label{fig.pulse_height_hidra} (a) Pulses on an SNSPD with incident light with different average numbers of photons. Pulse height is indicated. The count rate versus laser attenuation plot used to calibrate photon number is shown in the inset. (b) Pulse height versus photon number for the detector shown in (a). (c) Count rate for each of the five HiDRA SNSPDs used in this experiment. (d) Pulse height for each of the five HiDRA SNSPDs with count rates indicated in part (c). Average photon numbers for the first and last SNSPD are indicated.}
\end{figure}

\begin{figure}
\includegraphics[width=8.6cm]{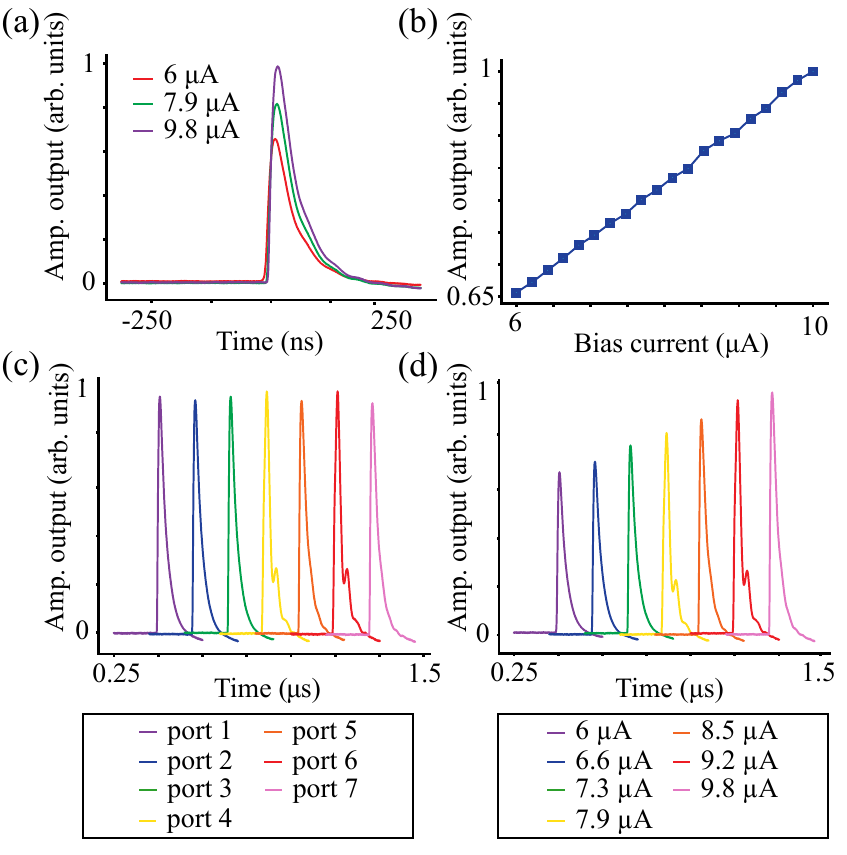}
\caption{\label{fig.yield_tree} (a) Pulses for three different values of bias current on a single detector. (b) The pulse height versus bias current for the detector in (a). (c) Pulses on each of seven detectors on an even tree device with a fixed bias current of 7.7 $\mu$A. (d) Pulses on the same seven detectors as in part (c) for seven different values of bias current. The x-axis separation of the pulses in parts (c) and (d) are for visibility purposes only.}
\end{figure}

In the superconducting optoelectronic neuromorphic platform proposed in Refs.\,\citenum{Shainline2019} and \citenum{Shainline2018}, communication is photonic, but computation is electronic. The signals communicated from each neuron to its synaptic connections are binary, few-photon pulses, meaning the information communicated is independent of the number of photons in the pulse. Only the timing of spikes is relevant. Synaptic weighting of the pulses is done entirely electronically. This communication scheme requires the detectors to be insensitive to the number of incident photons. We use the tree and HiDRA structures to test whether or not SNSPDs demonstrate this behavior by measuring their output electrical pulses when they are illuminated over a light intensity range of close to 50\,dB. At the low end of this range, multi-photon absorption events are very improbable, with less than $10^{-3}$ photons per pulse on average, while at the high end multi-photon events are likely, with greater than $10$ photons per pulse on average. The measurement is performed with an input laser with 50\,ps pulses operating at a pulse rate of 200\,kHz and wavelength of 1.57\,\textmu m. The photon number is changed via a calibrated variable fiber attenuator. For coherent light, the photon number should follow a Poisson distribution given by $p\left(n>0\right) = 1-e^{-\left< n_{ph} \right>}$. Since the laser pulse length is much shorter than the response time of the detector, at most one count is observed every time the photon number in a pulse incident on the detector is greater than zero. Therefore $p\left(n>0\right)$ can be approximated by the count rate on the detector $R_{\mathrm{det}}$ divided by the repetition rate of the laser, $R_{\mathrm{laser}}$. The average photon number in the pulse received on the detector should therefore be related to these quantities by $\left< n_{ph} \right> = -\log{\left(1-\frac{R_{\mathrm{det}}}{R_{\mathrm{laser}}}\right)}$. Since this is the mean photon number received on the detector, no assumptions so far have been made about the absolute value of the laser power. Next, we assume the laser power is proportional to the mean photon number, and therefore the plot of count rate versus attenuation can be used to extract the mean photon number incident upon the detector at each set laser power. Figure \ref{fig.pulse_height_hidra}(a) shows pulses from an SNSPD with waveguide-coupled light with mean photon numbers of 0.003 and 12. The difference is indistinguishable on this scale. The inset shows the count rate versus laser attenuation for that SNSPD, from which the mean photon numbers were calculated. Figure  \ref{fig.pulse_height_hidra}(b) shows pulse height versus mean photon number for the detector in part (a). One hundred pulses were averaged to obtain the pulse height at each photon number. The pulse height varies by less than 2\% over almost five orders of magnitude in incident photon flux, with no discernible dependence on the number of incident photons. Figure \ref{fig.pulse_height_hidra}(c) and (d) show the results using the same laser configuration for detectors on a HiDRA device. Part (c) shows the count rate versus port number on the HiDRA, while part (d) shows the pulse height for each of the five detectors for the count rates shown in part (c). The pulse height varies by less than $5\%$ over the HiDRA, despite using five different detectors with mean photon numbers varying from 0.02 to 6. Mean photon numbers were calculated for each detector in the HiDRA by the procedure described for the data in Fig. \ref{fig.pulse_height_hidra} (a) and (b). In the synaptic weighting scheme described in Ref.\,\citenum{Buckley2018}, the magnitude of the SNSPD current pulses are used as the weights. This pulse amplitude can be controlled with the bias current applied to the SNSPD. In this case, it is very simple to generate synapses with a bit depth of one by simply biasing the SNSPD or not. However, further weighting can also be achieved by varying the bias current within the detection range of the synapse. It has also been shown that neural networks that perform simple tasks can be designed with these simple integrate-and-fire neurons \cite{Buckley2018}. In Fig.\,\ref{fig.yield_tree}(a), pulses on an SNSPD for three different bias currents are shown. Figure \ref{fig.yield_tree}(b) plots the pulse height versus bias current for the detector over the available range of bias currents, indicating a range of 6-10\,\textmu A. The black line shows a linear fit to the curve. In Fig.\,\ref{fig.yield_tree}(c), pulses from seven detectors on an symmetrical tree are shown. This is a tree device with all seven detectors fabricated with the same geometric parameters. The pulses are separated on the $x$-axis (in time) for clarity, this separation has no physical meaning. Figure. \ref{fig.yield_tree}(d) shows the pulses on the same seven detectors with bias currents changed to give different pulse heights. In Ref.\,\cite{Buckley2018} the spiking neural network model assumed that bias currents could be varied between 5\,\textmu A and 15\,\textmu A (or set to zero). This is around twice the range that is observed here. It would be necessary to re-simulate networks with these values to ensure that this is still a viable method for controlling the weights. Calculating the bit depth in a quasi-analog system is challenging, especially before building the system and determining what constitutes the minimum reasonable change in weight value. This is a challenge for many emerging technologies, such as magnetic Josephson junctions, which have $>20,000$ internal states per \textmu m$^2$ and are therefore quasi-analog\,\cite{Schneider2018}, and spin-torque nano-oscillators\,\cite{Romera2018} where frequencies are tuned using an applied current. Memristor technologies have similar issues, with typically reported bit depths of 4-6 \cite{Hu2018}, with bit depths of up to eight reported for phase-change memory \cite{Ambrogio2018}. However, even in these technologies, arguments have also been made for operating in a binary mode \cite{Hirtzlin2020}, for which these synapses are well suited. Superconducting loop neurons\,\cite{Shainline2019} use SNSPDs to generate the electrical pulses and then Josephson junctions add current to an integrating loop in integer quantities of fluxons (although thermal noise will play a role, and further investigation is required). The maximum number of fluxons that can be added determines the bit depth, with an upper limit of around 10 bits. 

\section{Methods}
\label{sec.methods}

\subsection{Fabrication}
\label{sec.fab}

\begin{figure}
\includegraphics[width=8.6cm]{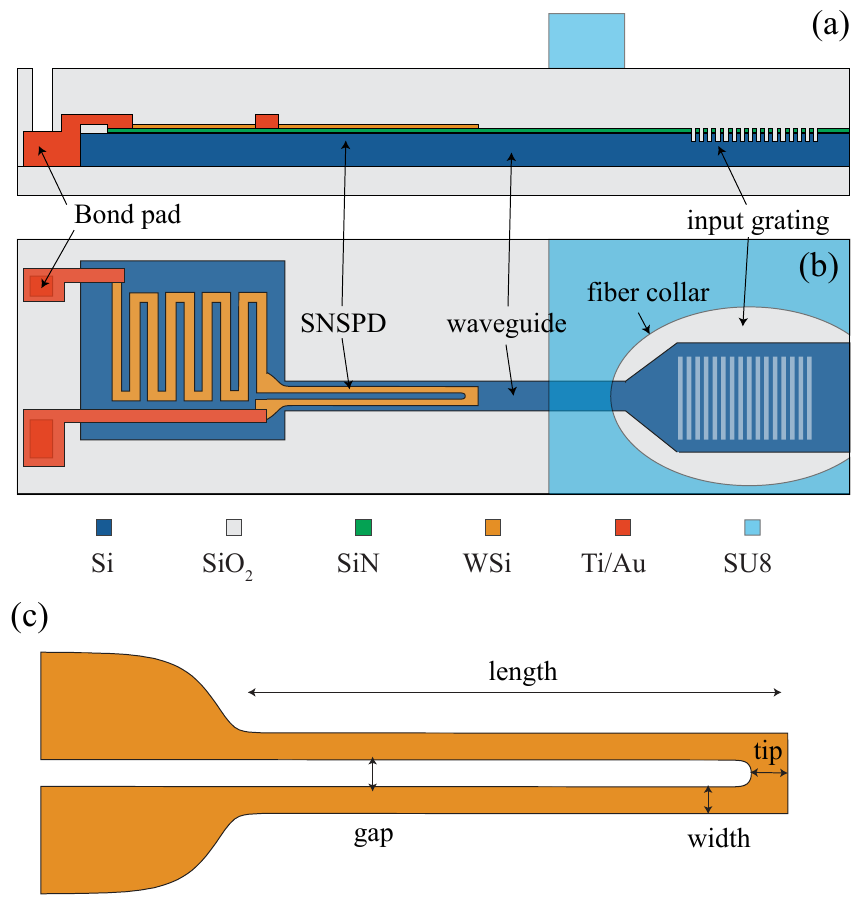}
\caption{\label{fig.fabrication} A schematic of the fabrication process used in the paper showing (a) a cross section and (b) a top view. (c) The geometry of the waveguide nanowire.}
\end{figure}

Waveguide-coupled SNSPDs were fabricated using a layer structure that builds toward the full superconducting-optoelectronic neuromorphic hardware fabrication process
 \cite{Buckley2020a}. Certain layers, while not strictly necessary for this demonstration, were included for compatibility purposes. These include the small pad metal layer, the waveguide 
 hardmask (similar to the dielectric spacer used in the amplifier fabrication process) and its associated via layer (all described below). The SNSPDs were made from WSi (shown in 
 orange in Fig. \ref{fig.fabrication} (a)), while the waveguides were silicon (blue). The substrate was a silicon-on-insulator (SOI) wafer with a 220\,nm device layer and a 3\,\textmu m 
 buried oxide layer. A 40\,nm SiN layer (green) was sputtered on the SOI to act as an etch stop for the WSi etch. Electron-beam and photolithography alignment marks were then etched 
 into the wafer. A 2\,nm/30\,nm/2\,nm Ti/Au/Ti layer was patterned via liftoff (red). This metal layer was used to make electrical contact to the SNSPDs. 

A 2.1\,nm W$_{0.64}$Si$_{0.36}$ film was then co-sputtered as described in Ref.\,\onlinecite{Chiles2020}. An amorphous silicon capping layer was then deposited without breaking vacuum to protect the thin WSi film. This WSi recipe has been demonstrated to allow saturated internal quantum efficiency for nanowires as wide as 1.2\,\textmu m at 1.55\,\textmu m \cite{Chiles2020}. The nanowires were patterned with  photolithography using a 365\,nm, i-line stepper and etched with Ar and SF$_6$ chemistry. 

A partial silicon etch was performed to create the grating used for coupling light from an optical fiber to the waveguides. This patterning was performed with electron-beam lithography. The etch depth was designed to be 50\,nm and measured at 65\,nm\,-\,70\,nm. A 100\,nm SiO$_2$ hardmask was then deposited using plasma-enhanced chemical vapor deposition (PECVD) and the waveguides were patterned and etched with electron-beam lithography using a positive tone resist with 2\,\textmu m clearance around the photonic devices. The remainder of the Si was then cleared out with photolithography. All silicon etches used SF$_6$ and C$_4$F$_8$ chemistry. 

Following the waveguide etch, vias were etched through the waveguide oxide hard mask to make electrical contact to the Au layer using CHF$_3$/O$_2$ chemistry. A wiring layer of 2\,nm/440\,nm/2\,nm Ti/Au/Ti was then deposited and patterned with liftoff. A 1.8\,\textmu m cladding oxide (white) was then deposited using PECVD. Openings to the wirebond pads and the ground plane were then etched using the same CHF$_3$/O$_2$ dry etch. Finally, a 50-\textmu m-thick SU8 packaging layer (light blue) was spun on and patterned with photolithography \cite{Shainline2017b}. 

Fig. \ref{fig.fabrication} shows (a) cross sectional and (b) top view diagrams of a fabricated device, indicating some of the main features. Optical microscope images and scanning electron-microscope images of fabricated devices are shown throughout this paper alongside measured data. The geometry of a typical SNSPD is shown in Fig. \ref{fig.fabrication} (c). The geometric parameters are indicated in the diagram. Length and width were varied in this paper to determine optimal device performance. 
\subsection{Photonic device design}
\label{sec.photonic_design}

\begin{figure}
\includegraphics[width=8.6cm]{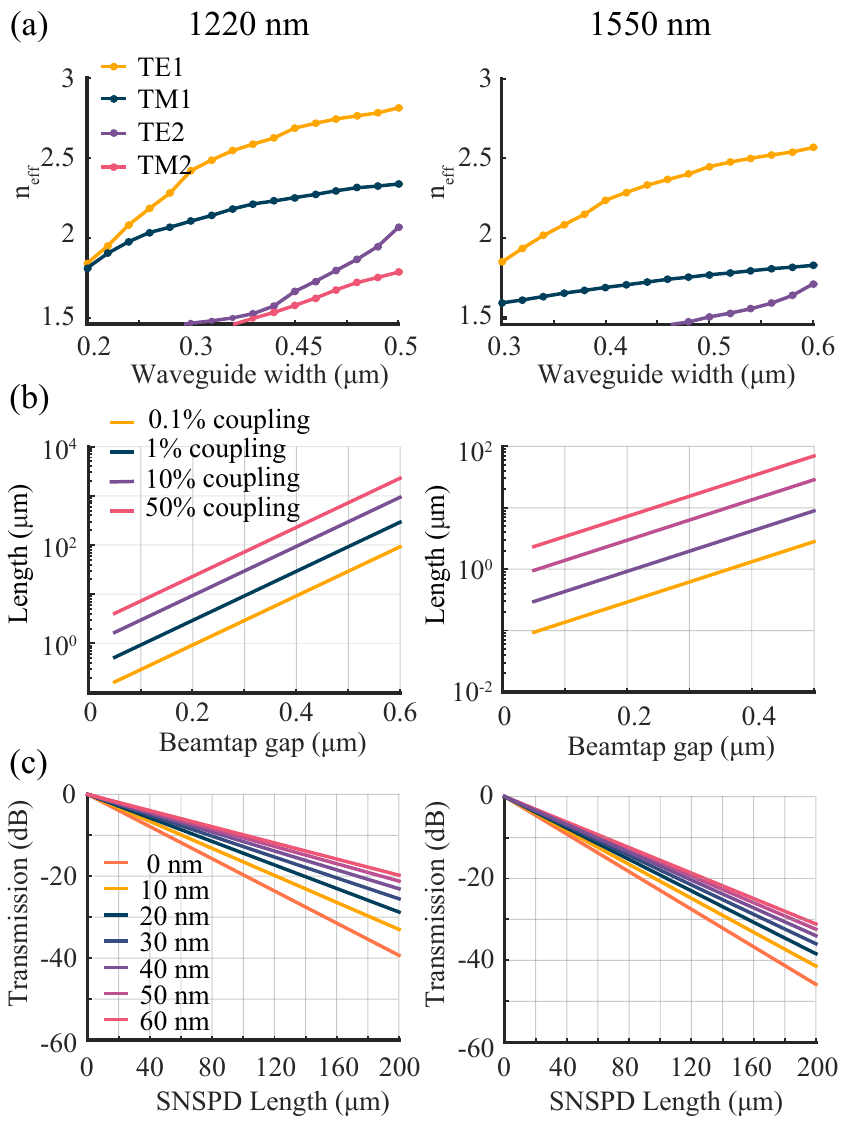}
\caption{\label{fig.design} (a) Simulated waveguide modes at 1.22\,\textmu m and 1.55\,\textmu m. (b) Simulated beamtap length versus gap for different values of splitting ratio at 1.22\,\textmu m and 1.55\,\textmu m. (c) Simulated transmission versus nanowire length for different values of the SiN spacer thickness.}
\end{figure}
The devices described in this paper required design of a number of components. These devices include waveguides, beamsplitters, beamtaps and SNSPD hairpins. Photonic devices were designed for operation at both 1.55\,\textmu m (standard telecommunications wavelength), and at 1.22\,\textmu m, where the proposed spiking neural networks \cite{Shainline2017a, Shainline2018,Shainline2019, Shainline2020} are designed to operate. Simulations of the photonic waveguides were performed using a finite-difference frequency-domain eigenmode solver. The effective index of waveguide modes in 220\,nm thick waveguides of different widths for 1.55\,\textmu m and 1.22\,\textmu m wavelengths are shown in Fig.\,\ref{fig.design} (a). Based on these simulations, single-mode waveguide widths were chosen to be 450\,nm (350\,nm) for waveguides designed for operation at 1.55\,\textmu m (1.22\,\textmu m). For branching-tree and high-dynamic range array structures, the beamtaps were simulated using the eigenmode method described in Ref.\,\onlinecite{Chrostowski}. The results of the simulations are shown in Fig.\,\ref{fig.design} (b). For the 1.55\,\textmu m beamtaps, the waveguide width was brought down to 400\,nm at the beamtap region. The beamsplitter gap widths and interaction lengths were chosen based on the plots in Fig\,\ref{fig.design}(b). The 1.55\,\textmu m beamsplitters were chosen based on the optimized design in Ref.\,\onlinecite{Zhang2013}. The 1.22\,\textmu m beamsplitters are multi-mode interferometers that were optimized in separate fabrication runs to minimize loss.

The waveguide-integrated nanowire geometry is shown in Fig.\,\ref{fig.fabrication}(c). We refer to this structure as a hairpin. Unless otherwise noted, the measured width of the nanowire was 500\,nm and the length was 200\,\textmu m. There was an 800\,nm gap between the wires in the hairpin, with a 400\,nm gap on either side. Waveguides were adiabatically tapered from their single-mode widths up to 2.6\,\textmu m over a 100\,\textmu m length to accommodate the SNSPD. The detectors also have a meandering section of 3\,\textmu m-wide wire for extra inductance to achieve a total value 1.25\,\textmu H. 

Nanowire absorption was calculated for a 500\,nm-wide nanowire with the geometry shown in Fig.\,\ref{fig.fabrication}(c). The real and imaginary parts of the refractive index of the WSi was measured and used to simulate the real and imaginary propagation constant in the nanowire on waveguide structure. The resulting absorption versus length at  1.22\,\textmu m and 1.55\,\textmu m are shown in Fig.\,\ref{fig.design} (c) for different thicknesses of SiN spacer. Based on the results of this plot in combination with concerns about ensuring electrical isolation of the superconducting layer, the SiN spacer was designed to have a 40\,nm thickness.

\subsection{Component characterization}
\label{sec.passive_characterization}
\begin{figure}
\includegraphics[width=8.6cm]{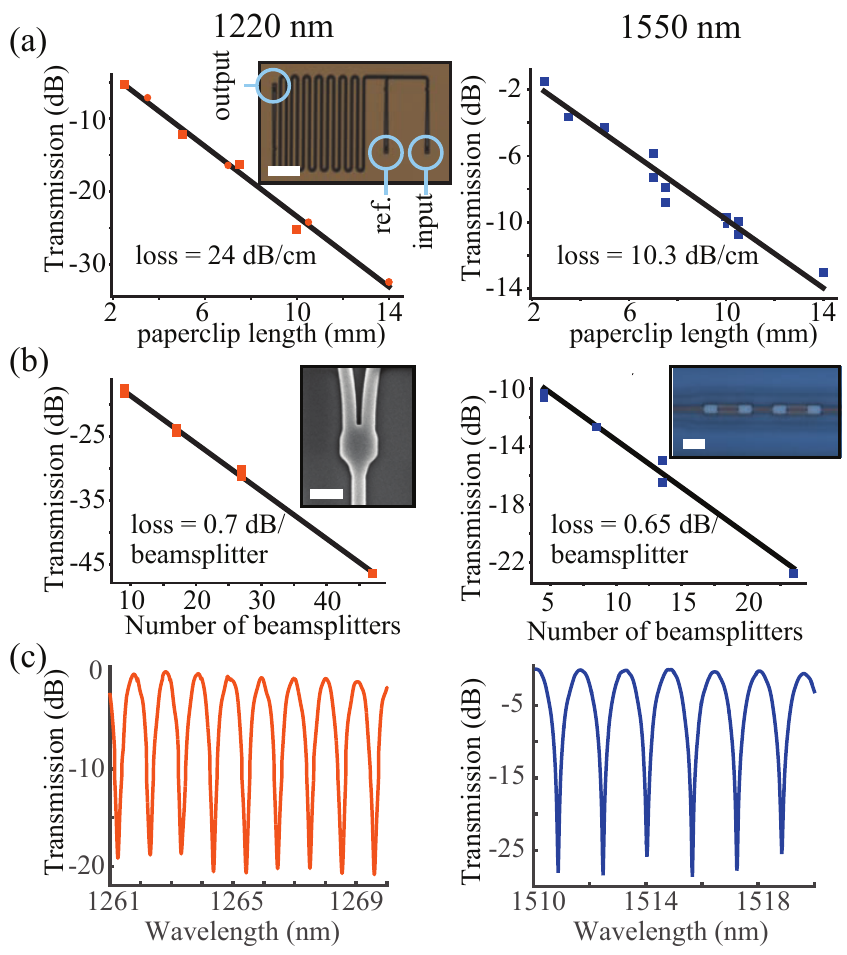}
\caption{\label{fig.wg_bs} (a) Measured waveguide loss at 1.22\,\textmu m and 1.55\,\textmu m with an optical microscope image of the structure used for loss measurement inset (scalebar = 200 \textmu m). (b) Measured beamsplitter loss at 1.22\,\textmu m and 1.55\,\textmu m with left inset showing an SEM of a beamsplitter (scalebar = 1 \textmu m) and right inset showing an optical microscope image of the structure used for measuring beamsplitter loss (scalebar = 5 \textmu m). (c) Measured transmission spectrum of Mach-Zehnder interferometer at 1.22\,\textmu m and 1.55\,\textmu m.}
\end{figure}
For every aspect of the integrated devices described in Sec.\,\ref{sec.cryo}, test structures were also fabricated to characterize performance and losses. These test structures are described below and in Fig.\,\ref{fig.wg_bs}. All of these test devices were fabricated on the same wafer as the devices with integrated nanowires.  Insertion losses of the grating couplers varied from 5-7\,dB at 1.55 \textmu m, and 8-10\,dB at 1.22 \textmu m.

The transmission of meandering waveguides of various lengths was used to measure the propagation loss of the waveguides (cutback method). The transmission as a function of waveguide length is shown in Fig.\,\ref{fig.wg_bs}(a). Each device has an input, a reference port, and an output grating, as shown in the optical microscope image in the inset. The intensity of the light from the reference and output gratings is measured with an infrared camera and compared for devices of different lengths to obtain the data in Fig.\,\ref{fig.wg_bs}(a). The measured losses are 24\,dB/cm at 1.22\,\textmu m and 10.2\,dB/cm at 1.55\,\textmu m. These losses can be improved with well-known processes \cite{Bojko2011}.

The loss per beamsplitter was measured using a similar technique to waveguide propagation loss. Transmission through sequences of various numbers of beamsplitters was measured. The data are shown in Fig.\,\ref{fig.wg_bs}(b), and the insets show a scanning electron micrograph of a single beamsplitter as well as a microscope image of a series of four beamsplitters. For this measurement, light was collected through a second fiber and detected with a photodiode. The loss of 0.7\,dB per beamsplitter is around double the insertion loss simulated \cite{Chrostowski}. To quantify the power splitting ratio of the beamsplitters, they were embedded in unbalanced Mach-Zehnder interferometers (MZIs). The extinction of such an interferometer can be used to bound the splitting ratio \cite{Chrostowski}. Transmission spectra of MZIs at 1.22\,\textmu m and 1.55\,\textmu m are shown in Fig.\,\ref{fig.wg_bs}(c). The extinction at 1.55\,\textmu m was nearly 30\,dB, while at 1260\,nm (measured at this wavelength due to the availability of an O-band laser) it was near 20\,dB. From these values we get approximate splitting ratios of 48\%/52\% at 1.55\,\textmu m and 45\%/55\% at 1260\,nm. The extinction values measured with the MZIs were similar to the noise floor observed with SNSPDs at those wavelengths (see Secs.\,\ref{sec.trees} and \ref{sec.hidras}). This result may indicate the beam splitting ratios were closer to 50\%/50\% than calculated from these measurements. 
 

\subsection{Cryogenic measurements}
Room-temperature and cryogenic measurements were made on devices comprising both integrated photonic and superconducting nanowire components. For cryogenic measurements, detector samples were wire-bonded and fiber-packaged by the method described in \cite{Shainline2017b}. The packaged samples were then cooled to 800\,mK in a closed cycle sorption pump ${}^4$He cryostat. Detectors were DC biased and measured through AC-coupled amplifiers. Light could either be coupled onto the chip via the packaged fibers, or via a separate fiber that flooded the entire chip uniformly with light. 

For each cryogenic measurement, a corresponding room-temperature measurement was also performed on a similar device with output gratings in place of single-photon detectors. For these room-temperature measurements, we used a technique that has been previously described for characterization of photonic structures in Ref.\,\citenum{Chiles2018}. With this technique, light is coupled from an optical fiber to an input grating, and the device is characterized by measuring the intensity emitted from output gratings at different points in the device on an infra-red camera. The room-temperature measurements are directly comparable to the cryogenic measurements. 


\subsection{High dynamic range room temperature camera measurements}
\label{sec.high_dynamic_range}
Parallel automated measurements of the room-temperature properties of passive photonic devices were made using a procedure similar to that described in Ref. \citenum{Chiles2018}. In brief, light was input through a single input grating. The light from the output gratings following the nanowires and reference paths is captured on an infrared-sensitive camera. Automated image processing was used to find the output grating locations and extract the relative intensities of light from these outputs. The green boxes in Fig.\,\ref{fig.tree_images}(d) indicate the locations that have been automatically assigned as the outputs. The intensity of the light after each nanowire is compared to the intensity from the associated reference grating to normalize out the effect of the grating response as well as propagation losses. The structures measured on this wafer have high dynamic range, with up to 30\,dB attenuation measured in the devices themselves. This poses an issue for the 8-bit camera used for characterization of the structures. We therefore used a calibrated attenuator to vary the input light and capture the light from the reference and measurement gratings over 50\,dB of attenuation. 

\subsection{Tree analysis}
\label{sec.tree_analysis}

\begin{figure}[h!]
\includegraphics[width=8.6cm]{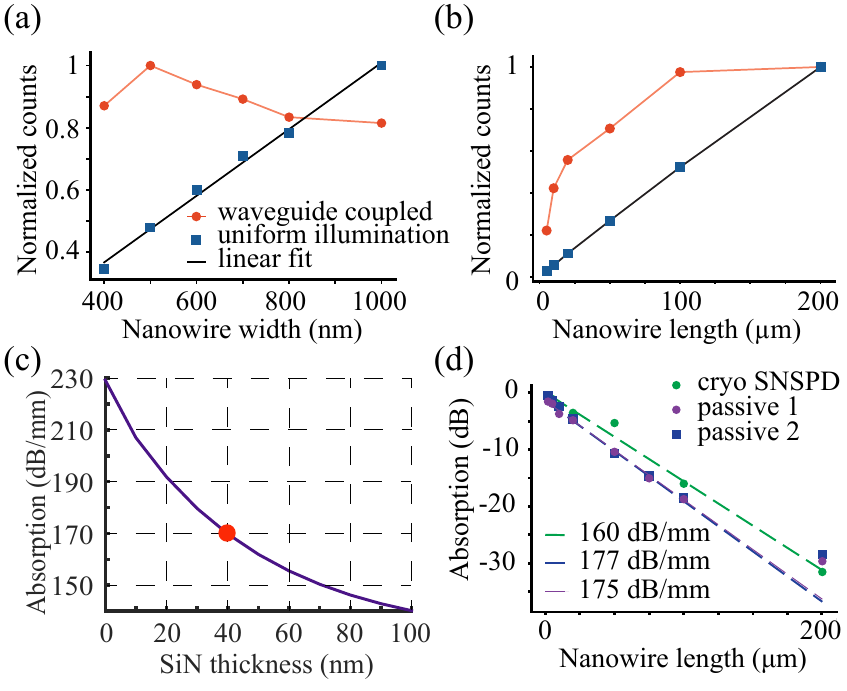}
\caption{\label{fig.tree_data_methods} (a) The height of the plateau versus the nanowire width for uniform illumination (blue squares) and waveguide coupled light (red circles). The count rate is linearly proportional to the nanowire width when the chip is uniformly illuminated, as shown by the linear fit (black line). (b) Count rate versus nanowire length for six different SNSPD lengths for waveguide-coupled and not-waveguide-coupled light. The count rate is linearly proportional to the SNSPD length when the detectors are uniformly illuminated, while the waveguide-coupled data saturates. (c) Simulation data for the absorption coefficient versus the SiN spacer thickness, extracted from the data in Fig. \ref{fig.design}\,(c). The red dot indicates the fabricated value. (d) Absorption versus nanowire length for cryogenic and passive devices measured at room temperature, showing absorption between 160\,dB/mm and 175\,dB/mm.}
\end{figure} 

In Fig.\,\ref{fig.tree_data_methods} (a) the plateau height (averaged value of the data points within the plateau region) is plotted versus wire width for the tree with nanowires of different widths. The blue squares show the result when the chip is uniformly illuminated. The black line is a linear fit, which is expected due to the fact that the detector area is proportional to the wire width. The red circles show the same analysis for the waveguide-coupled data. Since all of the nanowires absorb more than 99\% of the light, there should be no discernible difference between the values for the different wire widths. However, the data indicates a slight drop in efficiency for wider wires. The cause of this drop in efficiency may be the tip of the wire, which is approximately four times wider than the rest, absorbing some of the light without leading to detection events. This design is intended to avoid current crowding \cite{Clem2011}. For reference, a 4\,\textmu m-length corresponds to 14\% absorption according to the measured absorption values (simulated in Fig.\,\ref{fig.design} and measured in Fig.\,\ref{fig.tree_data_methods}(d)). It may be possible to narrow this tip length, although the cost may be a lower switching current and narrower plateau region. Interestingly, the narrowest wire is also less efficient, possibly due to edge roughness resulting from pushing the limits of the photolithography tool used in fabrication. 

Figure \ref{fig.tree_data_methods}(b) shows the height of the nanowire pleateau versus the nanowire length for the different length nanowires. Normalization is again performed against the highest-count-rate detector on that particular tree device. A linear dependence versus length (proportional to area) is again observed for the case of uniform illumination. The same nanowires show a saturating behavior with increasing length when the light is waveguide coupled, indicating that all light has been absorbed in the longest wires. We use this tree to quantify the absorption of the nanowires versus length. The expected absorption coefficient can be extracted from the simulations shown in Fig. \ref{fig.design}(c). A plot of simulated absorption coefficient versus SiN spacer thickness shown in Fig. \ref{fig.tree_data_methods}(c) with the value expected for the fabricated SiN thickness indicated by the red dot. Figure \ref{fig.tree_data_methods}(d) shows $10\cdot\log(A-R_{\mathrm{det}})$, where $R_{\mathrm{det}}$ is the SNSPD count rate, versus length for the value of $A$ that gave the best linear fit (green dashed line). The nanowire absorption per unit length calculated from the slope of this line is 160\,dB/mm.

We also measured the waveguide-integrated nanowire absorption at room temperature. Separate photonic trees were fabricated for this purpose. Such a tree is shown in Fig.\,\ref{fig.tree_images}(d). In these structures, output grating couplers were fabricated after each nanowire. Each waveguide-integrated nanowire is also accompanied by a reference waveguide with the same geometry but without a nanowire. Light is coupled through a fiber to the input grating. A branching tree splits the light evenly along 16 paths in this structure, with eight nanowires each accompanied by a reference waveguide, for a total of 16 output ports. The automated measurement technique is described in Methods Sec. \ref{sec.high_dynamic_range}.

The results of this measurement for the tree structure with varying nanowire length are compared to cryogenic measurements of counts per second in Fig.\,\ref{fig.tree_data_methods}(d), purple circles and blue squares (indicating measurements of two equivalent devices). The absorption in this case is directly measured (unlike the case of the cryogenic measurement, where the total light to be absorbed is a fit parameter). We find 175\,dB/mm attenuation (purple and blue lines) from the SNSPD hairpin at room temperature, compared with the cryogenic measurement of 160\,dB/mm. This level of agreement may be satisfactory for some applications, in which case it would be sufficient to measure the absorption of prototype nanowires with room temperature measurements. The simulations shown in the Methods give a value of 165\,dB/mm for the structures studied in the experiment, which is in good agreement with the experimental results.  

\subsection{HiDRA analysis}
\label{sec.hidra_analysis}

\begin{figure}[h!]
\includegraphics[width=8.6cm]{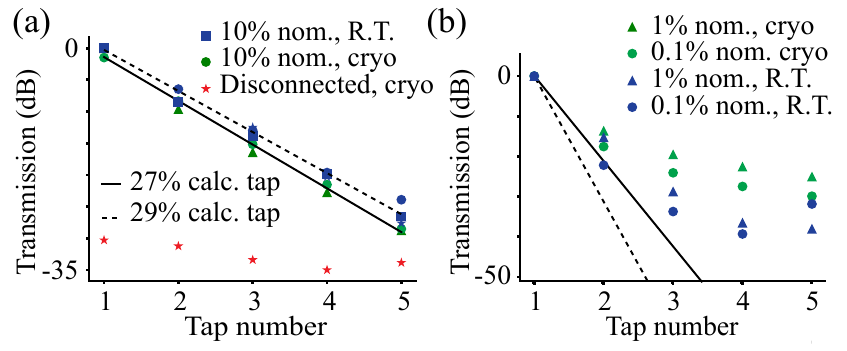}
\caption{\label{fig.hidra_analysis_methods} (a) Count rate versus port number for a HiDRA designed to have a 10\% beamtap, including measurements of two cryogenic detector devices (green markers), two room temperature devices (blue markers). The black lines show linear fits to the cryogenic (solid) and room temperature (dashed) data. The red stars show the result for a cryogenic device with intentionally disconnected detectors. (b) Count rate versus port number for HiDRA with different designed beamtap ratios for both cryogenic (green makers) and room-temperature (blue markers) operation. The scattered light prevents the dynamic range from exceeding 30\,dB. The black lines show how a 1\% (solid) and 0.1\% (dashed) measured device should perform, indicating the nominal values are far from realized.}
\end{figure} 

In the HiDRA device, after $n$ beamtaps, an SNSPD should receive $(1-f)f^{(n-1)}$ of the light. The waveguide loss must also be included, which is approximated by $e^{2 \alpha \cdot l (n - 1)}$, where $\alpha = \frac{\ln{10}}{20} L$, $L$ is the waveguide loss in dB/m, and $l$ is the length of each segment of the HiDRA. The total light intensity at the $n$th detector is therefore $(1-f)f^{(n-1)}e^{2 \alpha \cdot l (n - 1)}$.  

Figure \ref{fig.hidra_analysis_methods}(a) shows further analysis of the devices designed to have a 10\% beamtap. The log of detector count rate versus detector tap number $n$ is a straight line, indicating that the tap fraction is constant. The tap fraction can be calculated from the slope $m$ of this plot, as well as the known segment length $l$ and waveguide loss $L$ in dB/m. The tap fraction $f$ is then given by $f = 10^{\frac{m - Ll}{10}}$. We extract tap fractions of 0.26 and 0.28 from the two HiDRAs designed to have taps with $f=0.1$, with the mean value (solid black line) of 0.27. This deviation from the designed value is likely due to the the gaps coming out narrower than designed. This in combination with a slightly longer effective length due to the sine bends leading up to the beamtap could explain the discrepancy. A third device with intentionally disconnected waveguides was also measured (red stars). Disconnected HiDRAs were identical aside from a 180$^\circ$ waveguide turn and taper that couples light into free space, away from the SNSPD. The disconnected device shows a relatively constant level at around 30\,dB below the level of the first waveguide coupled detector on the connected HiDRA. This indicates that with our current fabrication process and packaging scheme there is a 30\,dB noise floor. Previous measurements of LEDs coupled to waveguides indicated a 40\,dB noise floor \cite{Buckley2017}. Both the low fiber-coupling efficiency and high waveguide scattering loss contribute to this high background light level.

Figure \ref{fig.hidra_analysis_methods}(a) also shows the transmission for each of the output ports of a passive, room-temperature HiDRA. From the plot of transmission versus attenuation, the beamtap ratio at room temperature is determined to be 0.29 for these devices, calculated from the slope of the dashed black line in the figure. This value agrees well with the cryogenic measurements described in the preceding paragraph. 

The results of cryogenic and room temperature measurements of HiDRA designed to have 1\% and 0.1\% beamtaps are shown in Fig. \ref{fig.hidra_data}(d). The solid black line indicates where the data for a cryogenic HiDRA with a 1\% tap ratio should fall, while the dashed black line indicates the same for a 0.1\% tap ratio device. The data clearly do not follow this trend. The scattered background light prevents such low tap ratios from being useful. The passive HiDRA with beamtap ratios designed to be 1\% (blue triangles) and 0.1\% (blue circles) indicate similar trends. We again find very good agreement with the cryogenic measurement, even observing the same roll off at 30\,dB noise floor.

\subsection{\label{apx.1220}Waveguide-integrated nanowire measurements at 1.22\,\textmu m}
\begin{figure}[h!]
\includegraphics[width=8.6cm]{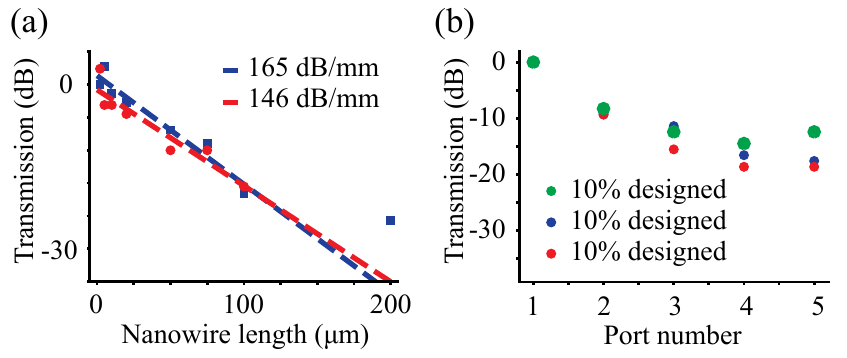}
\caption{\label{fig.1220nm} Measurements of (a) length trees and (b) HiDRA at room temperature at 1.22 \textmu m.}
\end{figure}
The values of absorption and dynamic range measured with the passive tree and HiDRA devices measured at 1.55\, \textmu m matched the active devices, and therefore we measured the 1.22\,\textmu m devices at room temperature without also performing cryogenic measurements. The measurement technique is the same as that described in the room-temperature measurements portions of Sections \ref{sec.trees} and \ref{sec.hidras} of the main text, but with a 1.22\,\textmu m input laser. We found that the measured absorption of the nanowires at 1.22\,\textmu m was similar to that measured at 1.55 \,\textmu m, giving 146\,dB/mm and 165\,dB/mm for two different (nominally identical) devices, as shown in Fig. \ref{fig.1220nm}(a). The large variation may be due to the fact that there is more scattering from the 1.22\,\textmu m waveguides due to edge roughness, which affects the accuracy of the camera intensity measurements.

The HiDRA devices designed to have 10\% beamtaps suffered from a very high background, with the intensity flattening out at around 20 dB, as shown in Fig. \ref{fig.1220nm}(b). This reduced extinction ratio is likely due to the very high waveguide scattering loss.

\section{Discussion}
\label{sec.discussion}
Realization of the hardware necessary for several forms of advanced cryogenic computing involves integrating devices in a scalable fabrication process. One of the main elements required for quantum and neural computing hardware involving photonic communication is high-yield, high-efficiency detectors. The work presented here introduces several tools leveraging integrated-photonic circuits for the characterization of waveguide-integrated SNSPDs, and makes several measurements important for synaptic operation. 

The integration of the detectors with photonic structures provides a powerful toolbox to rapidly assess large numbers of detectors with various design parameters in a context similar to that in which the detectors will be utilized. We have described the fabrication of waveguide-coupled SNSPDs that can be used as components in synapses in superconducting optoelectronic networks. We have demonstrated several integrated photonic structures that can be used to assess important properties of waveguide-integrated SNSPDs using a single fiber input, and we have compared room-temperature measurements with cryogenic measurements. We have demonstrated the effects of photon number and bias current on SNSPD pulse height, which are relevant to synaptic operation. Such measurement methods leveraging integrated-photonic circuits are likely to be indispensable in the maturation of advanced computing technologies using photons for communication.

A primary focus of this work has been on realizing SNSPDs that have high yield, enabling many such detectors to be integrated in large computational systems. To enable this, we have chosen to work with wide wires fabricated with photolithography rather than narrower wires commonly fabricated with electron-beam lithography.  In total throughout this paper, we have tested 49 waveguide-coupled SNSPDs, of which all 49 yielded with plateaus with $P>0.1$. These wider wires consume more area, which is problematic in large, integrated systems. However, we have found that elimination of even a single electron-beam lithography step greatly simplifies fabrication. This compromise will enable near-term demonstrations in academic and government lab cleanrooms, which is important for the scientific goal of building and testing superconducting optoelectronic neural networks. By simplifying the fabrication of the detectors to be compatible with our 365\,nm photolithography equipment, we have been able to fabricate large numbers of detectors with saturated internal efficiencies in a single device, something that has proved challenging in other materials \cite{Najafi2015,Dietrich2016,Gourgues2019}, in particular when integrated in complex photonic circuits \cite{Elshaari2020}. Ultimately, if such detectors prove valuable to advanced neural or quantum computing, the improved fabrication and readily achievable feature sizes available at a commercial foundry will only improve the performance demonstrated here, while reducing the area of each detector considerably. When fabricated in a commercial foundry, it may be beneficial to optimize processing for an entirely different material with a much higher T$_c$. This would allow the cryogenic complexity to be reduced by operating the entire platform above 4 K. Nevertheless, the techniques described in this work will be necessary for characterization of large numbers of SNSPDs in any material.

Since the inductance is set so that the $L/R$ time constant is sufficiently long to avoid latching, in this work large inductance meanders were added in series with the detection hairpins to bring the total value of the inductances $L$ above 1 \textmu\,H and achieve the correct $L/R$. This led to a large increase of the area of the devices. However, in a practical neuronal circuit, the inductance may be very different. In these measurements, SNSPD pulses were read out across a 50\,$\Omega$ transmission line.  In the context of synapses, the detectors will be in parallel with a much smaller resistance, and the inductance (and therefore the size) can be reduced while maintaining the same $L/R$ time constant for resetting. If the synapses of Ref.\,\cite{Buckley2018} are employed, the resistance affecting the SNSPD $L/R$ time constant is related to the integration time of the neuron. If the synapses of Refs.\,\cite{Shainline2018,Shainline2019} are employed, the normal-state resistance of the Josephson junctions (on the order of $1\,\Omega$) establishes a minimum functional value of SNSPD inductance, on the order of 100\,nH, or around a factor 10 smaller than in the present work.
 
The experiments presented here point to important challenges for the scaling of future technologies. The HiDRA device showed that in the present devices scattered light was significant, only 30\,dB below the measured signal. This finding is particularly relevant for photonic quantum computing, in which relatively strong pump lasers are used to generate entangled-photon pairs on chip, as well as for superconducting optoelectronic networks used for spiking neural systems in which many light sources must be integrated on a chip with many single-photon detectors. The scattered light from a neuron must be less than the waveguide coupled light pulses. Scattered light therefore puts an upper limit on the number of connections possible per neuron. This demonstrates that for most large-scale photonic systems incorporating on-chip detectors, some form of detector shielding will likely be necessary.

Finally, the synaptic measurements made in this paper confirm the utility of SNSPDs for the synaptic application, where the physics of the superconducting to normal phase transition is used as the binary signaling mechanism for neuromorphic computing. Future work will build on this to demonstrate complex synaptic circuits integrated with transmitters for integrate-and-fire functionality.
  
\begin{acknowledgments}
This is a contribution of NIST, an agency of the U.S. government, not subject to copyright.
\end{acknowledgments}

 \newcommand{\noop}[1]{}

\end{document}